\documentclass[a4paper,11pt,english]{article}
\usepackage[latin1]{inputenc}
\usepackage[english]{babel}
\usepackage{graphicx}

\usepackage{verbatim,times,bbm}
\usepackage{color}

\usepackage{amsmath,amssymb}

\usepackage[bindingoffset=0cm,left=2cm,top=3cm,bottom=3cm,right=2cm]{geometry} 

\newcommand{\chrho}{\check{\rho}}
\newcommand{\trho}{\tilde{\rho}}
\newcommand{\ima}{{\mathrm{i}\hspace{0.3mm}}}

\newcommand{\filta}{{L^\sista}}
\newcommand{\filtb}{{L^\sistb}}

\newcommand{\spa}{\hspace{-2mm}}
\newcommand{\ccn}{\|_{\mbox{\tiny $\mathrm{CCN}$}}}
\newcommand{\sub}{\|_{\mbox{\tiny $\mathrm{SUB}$}}}

\newcommand{\lhr}{\mathcal{L}_{\mathbb{R}}(\mathcal{H})}
\newcommand{\lhra}{\mathcal{L}_{\mathbb{R}}(\hilba)}
\newcommand{\lhrb}{\mathcal{L}_{\mathbb{R}}(\hilbb)}

\newcommand{\sista}{\mathsf{A}}
\newcommand{\sistb}{\mathsf{B}}
\newcommand{\hilba}{{\mathcal{H}_{\mbox{\tiny $\mathsf{A}$}}}}
\newcommand{\hilbb}{{\mathcal{H}_{\mbox{\tiny $\mathsf{B}$}}}}
\newcommand{\na}{{N_{\mbox{\tiny $\mathsf{A}$}}}}
\newcommand{\nb}{{N_{\mbox{\tiny $\mathsf{B}$}}}}
\newcommand{\naq}{{N_{\mbox{\tiny $\mathsf{A}$}}}^{\hspace{-1mm}2}}
\newcommand{\nbq}{{N_{\mbox{\tiny $\mathsf{B}$}}}^{\hspace{-1mm}2}}
\newcommand{\dens}{\mathcal{D}(\mathcal{H})}
\newcommand{\densa}{\mathcal{D}(\hilba)}
\newcommand{\densb}{\mathcal{D}(\hilbb)}
\newcommand{\lh}{\mathcal{L}(\mathcal{H})}
\newcommand{\lha}{\mathcal{L}(\hilba)}
\newcommand{\lhb}{\mathcal{L}(\hilbb)}
\newcommand{\tr}{\mathrm{tr}}
\newcommand{\dd}{{\mathsf{d}}}

\newcommand{\RR}{{\mathsf{R}}}
\newcommand{\rea}{{\mbox{\tiny $\mathrm{R}$}}}

\newcommand{\minor}{{\mbox{$\mathrm{M}$}}}

\newcommand{\hs}{{\mbox{\tiny $\mathrm{HS}$}}}

\newcommand{\mara}{\rho_\sista}
\newcommand{\marb}{\rho_\sistb}
\newcommand{\tra}{\mathrm{tr}_\mathsf{A}}
\newcommand{\trb}{\mathrm{tr}_\mathsf{B}}
\newcommand{\eau}{\mathfrak{E}_1^\sista}
\newcommand{\ebu}{\mathfrak{E}_1^\sistb}
\newcommand{\ean}{\mathfrak{E}_{\ennebis}^\sista}
\newcommand{\ebn}{\mathfrak{E}_{\ennebis}^\sistb}
\newcommand{\enne}{\mbox{\scriptsize $\mathsf{n}$}}
\newcommand{\ennebis}{\mbox{\footnotesize $\mathsf{n}$}}

\newcommand{\op}{\mathfrak{E}_{\un\sosp\enne}^{\sista,\sistb}(\rho)}
\newcommand{\map}{\mathfrak{E}_{\un\sosp\enne}^{\sista,\sistb}}
\newcommand{\un}{\mbox{\tiny $1$}}
\newcommand{\undu}{\mbox{\tiny $1,\hspace{-0.4mm}2$}}
\newcommand{\sosp}{\mbox{\tiny $\hspace{-0.2mm},\hspace{-0.4mm}...,\hspace{0.1mm}$}}
\newcommand{\eak}{\mathfrak{E}_k^\sista}
\newcommand{\ebk}{\mathfrak{E}_k^\sistb}
\newcommand{\eal}{\mathfrak{E}_l^\sista}
\newcommand{\ebl}{\mathfrak{E}_{l}^\sistb}

\newcommand{\abi}{(\mbox{\footnotesize $\mathsf{A}\rightarrow\mathsf{B}$})}

\newcommand{\eadu}{\mathfrak{E}_2^\sista}
\newcommand{\ebdu}{\mathfrak{E}_2^\sistb}

\newcommand{\ope}{\mathfrak{E}_{\undu}^{\sista,\sistb}(\rho)}

\newcommand{\coco}{\mathrm{c.c.}}
\newcommand{\supa}{\hspace{0.6mm}\epsilon^\sista}
\newcommand{\supb}{\hspace{0.6mm}\epsilon^\sistb}

\newcommand{\hsi}{\hat{\sigma}}

\newtheorem{definition}{Definition}[section]

\newtheorem{rmk}[definition]{Remark}
\newtheorem{theorem}[definition]{Theorem}

\begin{document}

\title{On the relation between Schmidt coefficients and entanglement}

\author{Paolo Aniello \\
Dipartimento di Scienze Fisiche dell'Universit\`a di Napoli
``Federico II'' \& INFN -- Sezione di Napoli \& \\ Facolt\`a di
Scienze Biotecnologiche, Universit\`a di Napoli ``Federico II'', via
Cintia I-80126 Napoli, Italy
\and Cosmo Lupo\\
Dipartimento di Fisica, Universit\`a di Camerino, I-62032 Camerino,
Italy}


\maketitle

\begin{abstract}
We consider the Schmidt decomposition of a bipartite density
operator induced by the Hilbert-Schmidt scalar product, and we study
the relation between the Schmidt coefficients and entanglement.
First, we define the Schmidt equivalence classes of bipartite
states. Each class consists of all the density operators (in a given
bipartite Hilbert space) sharing the same set of Schmidt
coefficients. Next, we review the role played by the Schmidt
coefficients with respect to the separability criterion known as the
`realignment' or `computable cross norm' criterion; in particular,
we highlight the fact that this criterion relies only on the Schmidt
equivalence class of a state. Then, the relevance
--- with regard to the characterization of entanglement ---
of the `symmetric polynomials' in the Schmidt coefficients and a new
family of separability criteria that generalize the realignment
criterion are discussed. Various interesting open problems are
proposed.
\end{abstract}


\section{Introduction}
\label{intro}

The study of the structure and of the properties of the convex body
of states of a quantum system is still nowadays a source of
intriguing problems and an active area of research, especially in
the case of quantum systems with a finite-dimensional Hilbert
space~{\cite{Bengtsson}}. A great impulse to this area of research
has been impressed by the new applications related to quantum
information science~{\cite{Nielsen}}. In particular, the interest of
many scientists is focused on the study of bipartite (and
multipartite) quantum systems, in which the typical quantum
phenomenon known as \emph{entanglement} can be present and has
remarkable consequences. Let us recall~{\cite{Werner}} that a state
$\rho$ of a bipartite quantum system is said to be \emph{separable}
if it can be written as a convex sum of product states
\begin{equation}
\rho = \sum_k p_k \ \rho_k^\sista \otimes \rho_k^\sistb,\ \ \
p_k>0,\ \ \ \sum_k p_k =1;
\end{equation}
otherwise, it is said to be \emph{entangled}. Clearly, an important
and difficult task is to determine whether a certain quantum state
is separable or not. To accomplish this task several necessary
and/or sufficient criteria have been formulated and discussed in
literature; however, no universal --- i.e.\ holding for any
dimension of the local Hilbert spaces --- necessary-sufficient
criterion that is practically implementable is known so
far~{\cite{Bengtsson,Horodecki}}.

In the following, we will consider the case of a bipartite quantum
system with a finite-dimensional carrier Hilbert space $\mathcal{H}
= \hilba \otimes \hilbb$, with $\hilba \cong \mathbb{C}^\na$ and
$\hilbb \cong \mathbb{C}^\nb$, $\na,\nb\ge2 $. Observe that the
states of the local subsystems $\sista$ and $\sistb$ are
characterized, respectively, by $\naq -1$ and $\nbq -1$ real
parameters (being unit-trace positive selfadjoint operators), while
the states of the global system are characterized by $\naq\nbq -1$
real parameters; this fact is at the very root of the quantum
entanglement. It is then not surprising that practically
implementable separability criteria can be obtained by considering a
convenient parametrization of the set of states of the composite
system.

Consider, for instance, the Fano parametrization~{\cite{Fano}} of a
$2\otimes 2$ bipartite state --- i.e.
\begin{equation}
\rho = \frac{1}{4}\left(
\mathbb{I}_2^\sista\otimes\mathbb{I}_2^\sistb + x_a\,
\sigma_a^\sista\otimes\mathbb{I}_2^\sistb + y_\alpha\,
\mathbb{I}_2^\sista\otimes\sigma_\alpha^\sistb + \xi_{a\alpha}\,
\sigma_a^\sista\otimes\sigma_\alpha^\sistb \right).
\end{equation}
(sum over the repeated indices being understood), where
$\{\sigma_a\}_{a=1}^3$ are the Pauli matrices --- or its
$\na\otimes\nb$ generalization in terms of the (traceless
selfadjoint) `Gell-Mann matrices' $\left\{ \Lambda_a^\sista |\
a=1,\ldots,\naq-1\right\}$ and $\left\{ \Lambda_\alpha^\sistb |\
\alpha=1,\ldots ,\nbq-1\right\}$:
\begin{equation} \label{blok}
\rho = \frac{1}{\na\nb}\left(
\mathbb{I}_\na^\sista\otimes\mathbb{I}_\nb^\sistb + x_a\,
\Lambda_a^\sista\otimes\mathbb{I}_\nb^\sistb + y_\alpha\,
\mathbb{I}_\na^\sista\otimes\Lambda_\alpha^\sistb + \xi_{a\alpha}\,
\Lambda_a^\sista\otimes\Lambda_\alpha^\sistb \right).
\end{equation}
The coefficients $\{x_a\}$, $\{y_\alpha\}$, $\{\xi_{a\alpha}\}$ ---
that are subjected to suitable constraints which guarantee the
positivity of the density matrix~{\cite{constr}} (while the
normalization of the trace is ensured by the form of the
parametrization itself)
--- turn out to be relevant for characterizing the entanglement
of the state $\rho$. In particular, the matrix formed by the
coefficients $\{\xi_{a\alpha}\}$ has been considered for obtaining
new separability criteria, see~{\cite{Devicente}}.

A natural generalization of decomposition~{(\ref{blok}}) is achieved
by considering `local bases of observables' (or `local orthogonal
observables'~{\cite{loo}})
\begin{equation} \label{locbases}
\left\{ F_a^\sista\colon\hilba\rightarrow\hilba |\
a=1,\ldots,\naq\right\},\ \ \ \left\{
F_\alpha^\sistb\colon\hilbb\rightarrow\hilbb |\
\alpha=1,\ldots,\nbq\right\}
\end{equation}
in the `local Hilbert-Schmidt spaces', i.e.\ orthonormal bases in
the real Hilbert spaces of selfadjoint linear operators in $\hilba$,
$\hilbb$ (endowed with the real Hilbert-Schmidt scalar product):
\begin{eqnarray}
\langle F_a^\sista, F_b^\sista \rangle_\hs \spa & = & \spa
\tr\Big(F_a^\sista\, F_b^\sista\Big)\! = \delta_{ab}, \\
\langle F_\alpha^\sistb, F_\beta^\sistb\rangle_\hs \spa & = & \spa
\tr\Big(F_\alpha^\sistb\,
F_\beta^\sistb\Big)\!=\delta_{\alpha\beta}.
\end{eqnarray}
Generalizing the expression~{(\ref{blok}}), one can now write
\begin{equation}\label{loo_dec}
\rho = \tilde{\rho}_{a\alpha}\, F_a^\sista \otimes F_\alpha^\sistb,
\end{equation}
where:
\begin{equation}\label{loo_mat}
\tilde{\rho}_{a\alpha} = \langle F_a^\sista\otimes F_\alpha^\sistb ,
\rho \rangle_\hs = \tr\Big( F_a^\sista\otimes F_\alpha^\sistb\, \rho
\Big).
\end{equation}
Note that the $\naq\times\nbq$ matrix of coefficients
$\tilde{\rho}\equiv\left[\tilde{\rho}_{a\alpha}\right]$ is formed by
--- at least in principle --- measurable quantities; namely, the
expectation values of (local) quantum observables.

Even more generally, one can relax the assumption that the local
bases~{(\ref{locbases})} consist of observables; i.e.\ one can
assume that they are generic orthonormal bases in the \emph{complex}
vector spaces of linear operators
--- in $\hilba$ and $\hilbb$, respectively --- endowed with the
complex Hilbert-Schmidt scalar product.

The matrix of coefficients $\tilde{\rho}$ has recently been
considered in connection with the separability problem, see
refs.~{\cite{Aniello,Lupo}}. In particular, it turns out that the
\emph{singular values}~{\cite{Horn}} of this matrix coincide with
the (mixed state) \emph{Schmidt coefficients} of $\rho$ (see below).
Moreover, the sum of these (non-negative) coefficients is equal to
the \emph{computable cross norm} of the state $\rho$, a quantity
which is at the base of the remarkable separability criterion known
as the `realignment' or `computable cross norm'
criterion~{\cite{ChenWu,Rudolph,Rudolph-bis}}.

The aim of the present contribution is to investigate the role
played by the Schmidt coefficients in the characterization of the
bipartite entanglement. There is a very sound reason for being
interested in such a question: the Schmidt coefficients have a
transparent mathematical meaning and are easily computable from
physically measurable quantities (actually we will see that they
are, in principle, physically measurable quantities themselves). We
will also indicate some interesting open problems.

The paper is organized as follows. In Sect.~{\ref{sec:classes}}, we
recall the notions of `Schmidt decomposition', of `Schmidt
coefficients' and of `Schmidt equivalence class' of bipartite
states. In Sects.~{\ref{sec:ent}} and~{\ref{sec:ent-bis}, we will
try to highlight the relation between entanglement and the Schmidt
coefficients. In particular, in Sect.~{\ref{sec:ent}}, we will
consider the link between entanglement and the sum of the Schmidt
coefficients of a bipartite state; while, in
Sect.~{\ref{sec:ent-bis}}, we will argue that a stricter
characterization of entanglement can be achieved by considering the
whole set of the `symmetric polynomials' in the Schmidt coefficients
rather than the sum of the Schmidt coefficients alone (which is the
symmetric polynomial of degree one). The results outlined in
Sect.~{\ref{sec:ent-bis}}, by virtue of the well known relation
between quantum states and quantum maps, allow to to obtain
--- see Sect.~{\ref{duality}} --- an interesting characterization of
entanglement breaking quantum channels. In Sect.~{\ref{sec:ineq}}, a
recently discovered new class of separability criteria ---
see~{\cite{Aniello}} --- is described. This class --- that contains
as particular cases various separability criteria already present in
the literature, including the realignment criterion --- relies on
the sum of the Schmidt coefficients of a `transformed operator'
suitably associated with the bipartite state whose separability is
under scrutiny. Eventually, in Sect.~{\ref{conclusions}}, a few
conclusion are drawn.


\section{Schmidt coefficients and Schmidt equivalence
classes of bipartite states} \label{sec:classes}

The \emph{Schmidt decomposition} plays a central role in the
characterization of pure state entanglement in the bipartite
setting~{\cite{Vidal}}. Let us recall (see~{\cite{Peres}}) that any
state vector $|\psi\rangle$, belonging to the bipartite Hilbert
space $\mathcal{H}=\hilba\otimes\hilbb$, can always be decomposed as
\begin{equation} \label{pst}
|\psi\rangle=\sum_{k=1}^\delta \lambda_k\,
|\psi_k^\sista\rangle\otimes|\psi_k^\sistb\rangle,\ \ \ \lambda_k\ge
0,
\end{equation}
where the vectors in the decomposition are tensor products of local
orthonormal systems ---
$\langle\psi_h^\sista|\psi_k^\sista\rangle=\langle\psi_h^\sistb|\psi_k^\sistb\rangle
= \delta_{hk}$ --- and $\delta \equiv \min\{\na,\nb\}$. This
expression, called `the Schmidt decomposition of the state vector
$|\psi\rangle$' (representative of a pure state), is non-unique,
while the \emph{Schmidt coefficients} $\{\lambda_k\}_{k=1}^{\delta}$
are uniquely determined (up to re-ordering). The set of the Schmidt
coefficients associated with a pure state completely characterizes
the separability of this state; accordingly, any measure of
entanglement on pure states can be written as a function of the
Schmidt coefficients~{\cite{Vidal}}. Unfortunately, this remarkable
property is lost in the more general context of mixed states. In
order to clarify this point, one has to extend the notion of Schmidt
coefficients to bipartite \emph{mixed} states. We will follow the
approach of refs.~{\cite{Aniello,Lupo}}.

Let us consider, then, the convex body $\dens$ of states of a
quantum system,\footnote{Since $\dim(\mathcal{H})<\infty$, by
Gleason's theorem we can identify $\dens$ with the set of density
operators (unit trace positive linear operators) in $\mathcal{H}$.}
with carrier Hilbert space $\mathcal{H}=\hilba\otimes\hilbb$. The
set $\dens$ can be regarded as a convex subset of the complex vector
space $\lh$ of linear operators in $\mathcal{H}$. The vector space
$\lh$ becomes a Hilbert space if endowed with the (complex)
Hilbert-Schmidt scalar product:
\begin{equation}
\langle A , B \rangle_\hs := \tr\big(A^\dag B\big),\ \ \ A,B\in\lh.
\end{equation}
We will denote by $\|\cdot\|_\hs$ the associated norm:
\begin{equation}
\| A \|_\hs := \sqrt{\tr\left(A^\dag A\right)}.
\end{equation}

Hence, given a density operator $\rho \in
\mathcal{D}(\hilba\otimes\hilbb)\subset\lh=\lha\otimes\lhb$, there
always exist local orthonormal systems $\{ E_a^\sista\}_{a=1}^\dd$,
$\{E_b^\sistb\}_{b=1}^\dd$
--- $\dd \equiv \min\{ \naq, \nbq\}=\delta^2$ ---
such that the density matrix $\rho$ admits a decomposition of the
form
\begin{equation}\label{S_dec}
\rho = \sum_{a=1}^\dd \mu_a\, E_a^\sista \otimes E_a^\sistb,
\end{equation}
where the coefficients $\{\mu_a\}_{a=1}^\dd$ are non-negative. We
stress that the the set of coefficients $\{ \mu_a\}_{a=1}^\dd$ is
uniquely determined, and these coefficients are the (mixed state)
Schmidt coefficients of the density operator $\rho$. In the case of
a pure state $\rho=|\psi\rangle\langle\psi|$ (now realized as a
$1$-dimensional projector), they are related in a simple way to the
state vector Schmidt coefficients of decomposition~{(\ref{pst})},
i.e.\ $\{ \mu_a
\}_{a=1}^\dd=\{\lambda_h\lambda_k\}_{h,k=1}^{\delta}$. One can
easily check that a state $\rho\in \dens$ is a separable pure state
--- $\rho=|\psi^\sista\rangle\langle\psi^\sista\hspace{-0.3mm}|\otimes|\psi^\sistb\rangle\langle\psi^\sistb
\hspace{-0.3mm}|$
--- if and only if the set of its Schmidt coefficients is of the form
$\{ \mu_a \}_{a=1}^\dd =\{1,0,0,\ldots\}$.

\begin{rmk} \label{rimarca}
It is clear that a Schmidt decomposition of the form~{(\ref{S_dec})}
can actually be written
for any operator belonging to $\lh$.\\
Note that in decomposition~{(\ref{S_dec})} (like in~{(\ref{pst})})
we allow the possibility that some Schmidt coefficients are zero,
while, in the literature, usually only non-zero Schmidt coefficients
are taken into account. Hence, according to our convention, in
general the set of the Schmidt coefficients $\{\mu_a\}_{a=1}^\dd$
may contain both zeroes and repeated elements. The number of nonzero
coefficients in the set $\{\mu_a\}_{a=1}^\dd$ is called the
\emph{Schmidt rank} of $\rho$, and will
be denoted by $\mathsf{R}$.\\
Usually, the Schmidt coefficients are arranged in non-increasing
order, but any particular ordering is irrelevant for our purposes.
\end{rmk}

It is easy to verify that the Schmidt coefficients of $\rho\in\dens$
coincide with the singular values of the matrix $\tilde{\rho}$
--- see~{(\ref{loo_mat})}--- for any couple of local orthonormal bases in
$\lha$ and $\lhb$. Since we can regard $\dens$ also as a convex
subset of the \emph{real} Hilbert space $\lhr=\lhra\otimes\lhrb$ of
\emph{selfadjoint} linear operators in $\mathcal{H}$ (endowed with
the real Hilbert-Schmidt scalar product), then there always exists a
Schmidt decomposition of $\rho$ of the form~{(\ref{S_dec})} where
the orthonormal systems $\{ E_a^\sista\}_{a=1}^\dd$ and
$\{E_b^\sistb\}_{b=1}^\dd$ consist of \emph{observables}. Hence, not
only the Schmidt coefficients are easily computable from physically
measurable quantities (the entries of the matrix $\trho$, associated
with given local bases of observables), but they are physically
measurable quantities themselves:\footnote{Note, however, that the
observables $\{ E_a^\sista\}_{a=1}^\dd$ and
$\{E_b^\sistb\}_{b=1}^\dd$ will depend on the state $\rho$.}
\begin{equation}
\mu_a=\tr\hspace{-0.5mm}\left(E_a^\sista \otimes
E_a^\sistb\,\rho\right).
\end{equation}

We have already mentioned the fact that the set of the Schmidt
coefficients of a bipartite state is uniquely determined by this
state; on the other hand, one can show that a set of Schmidt
coefficients does not identify a unique quantum state (or, more
generally, a unique linear operator). As it turns out that there is
a relation between the Schmidt coefficients and
entanglement~{\cite{Aniello,Lupo}}, it is worth wondering to what
extent these coefficients alone are able to witness the presence of
entanglement and, in this spirit, introducing a suitable `Schmidt
equivalence relation' (see~{\cite{Lupo}}).
\begin{definition}
We say that two bipartite density operators belonging to $\dens$ are
\emph{Schmidt equivalent} if they share the same set of Schmidt
coefficients.
\end{definition}

Observe that for any $\rho\in\dens$ --- since $\sum_{a=1}^{\dd}
\mu_a^2 =\langle\rho,\rho\rangle_\hs=\tr\big(\rho^2\big)$ --- we
have:
\begin{equation}
\sum_{a=1}^{\dd} \mu_a^2 =1\ \ \ \Leftrightarrow\ \ \ \rho\ \
\mbox{pure state};\ \ \ \ \ \sum_{a=1}^{\dd} \mu_a^2 <1\ \ \
\Leftrightarrow\ \ \ \rho\ \ \mbox{non-pure state}.
\end{equation}
Therefore, the Schmidt equivalence relation induces in $\dens$ two
kinds of equivalence classes: the \emph{pure state Schmidt
equivalence classes} (composed by pure states only) and the
\emph{non-pure state Schmidt equivalence classes}. The first kind of
equivalence classes can be easily characterized~{\cite{pure}}, while
a `complete' characterization of the second kind of classes seems to
be a difficult problem.

The pure state Schmidt equivalence classes are known to be smooth
manifolds~{\cite{pure}}. Whether this result can be extended to all
the Schmidt equivalence classes of bipartite states is an
interesting open problem. The technical difficulty in solving this
problem consists in the difficulty of verifying whether there exist
Lie groups acting smoothly and transitively on each Schmidt
equivalence class, see~{\cite{Lupo}}. In any case, it would be
interesting to achieve a more complete understanding of the
structure and the properties of the Schmidt equivalence classes of
bipartite states.


\section{Entanglement and the sum of the Schmidt coefficients}
\label{sec:ent}

In this section, we will consider the role played by the Schmidt
coefficients in the detection of entanglement. We will first focus
our attention on the so-called `computable cross norm' (CCN), and
then we will discuss its relation with the `realignment criterion'
(RC) for separability. Our exposition will be rather concise and
will not include proofs; the reader interested in the details may
consult refs.~{\cite{Aniello,ChenWu,Rudolph,Rudolph-bis}.

The CCN on $\lh=\lha\otimes\lhb$ is a norm $\|\cdot\ccn\colon
\lh\rightarrow \mathbb{R}^+$ defined as follows:
\begin{eqnarray}
\|C\ccn \spa & := & \spa \inf \Big\{ \sum_{k=1}^n \|A_k\|_\hs \,
\|B_k\|_\hs\colon\ C = \sum_{k=1}^n A_k\otimes B_k,
\nonumber\\
& & \spa \mbox{with}\; \{A_k\}_{k=1}^n\subset\lha,
\{B_k\}_{k=1}^n\subset\lhb,\ n\in\mathbb{N}\Big\},
\end{eqnarray}
for any $C\in\lh$. It is possible to show that it is a \emph{cross
norm} on $\lh$, namely,
\begin{equation}
\|A\otimes B\ccn= \|A\|_\hs\hspace{0.3mm} \|B\|_\hs,\ \ \ \forall
A\in\lha ,\ \forall B\in\lhb.
\end{equation}
What is the relation between the CCN and the Schmidt coefficients is
not immediately evident. However, it is possible to prove the
following fact. Let $C$ be an operator in $\lh$
--- in particular, a bipartite state $\rho\equiv C\in\dens$ --- and let $\{\mu_a\}_{a=1}^\dd$
be the set of its Schmidt coefficients; then, the following relation
holds:
\begin{equation}
\| C \ccn = \sum_{a=1}^\dd \mu_a.
\end{equation}
But, as already observed, the Schmidt coefficients of $C$ coincide
with the singular values of the matrix of coefficients of the
decomposition of $C$ with respect to any product orthonormal basis
in the tensor product Hilbert space $\lh=\lha\otimes\lhb$. Hence,
the CCN is easily computable, \emph{computable} cross norm indeed!
Explicitly, it turns out that the sum of the Schmidt coefficients of
$\rho$ coincides with the trace norm of the matrix $\tilde{\rho}$
--- see~{(\ref{loo_mat})} --- for any choice of the local bases $\{
F_a^\sista | \ a=1,\ldots,\naq\}$, $\{ F_\alpha^\sistb |\
\alpha=1,\ldots,\nbq\}$ in $\lha$ and $\lhb$, respectively (here we
are not assuming that these orthonormal bases consist of
observables):
\begin{equation}
\| \rho \ccn = \sum_{a=1}^\dd \mu_a = \| \tilde{\rho} \|_\tr ,
\end{equation}
where
\begin{equation}
\| \tilde{\rho} \|_\tr := \tr \Big( \hspace{-0.6mm}
\sqrt{\trho^\dag\hspace{0.3mm} \trho} \hspace{0.6mm} \Big),
\end{equation}

We now come to the RC. This criterion admits two formulations.
According to its more popular formulation~{\cite{ChenWu}} --- which
explains the term `realignment criterion' --- this separability
criterion is stated as follows (we will be rather sketchy; for the
details, see~{\cite{Aniello}}). Let $\rho\in\dens$ be a bipartite
state and let $\chrho$ be the $\na\nb\times\na\nb$ representative
matrix of this operator with respect to any product orthonormal
basis in the Hilbert space $\mathcal{H}=\hilba\otimes\hilbb$. One
can associate with the matrix $\chrho$ a \emph{realigned matrix}
$\chrho^\rea$ (which is a $\naq\times\nbq$ matrix obtained from
$\chrho$ via a `reshuffling' operation, a suitable re-arrangement of
the entries of $\chrho$), and it turns out that the following
implication holds:
\begin{equation}\label{RC0}
\rho \ \ \mbox{separable} \ \ \Rightarrow \ \ \| \chrho^\rea
\|_{\tr} \leq 1 .
\end{equation}

One can show that the realigned matrix $\chrho^\rea$ coincides with
the matrix $\tilde{\rho}$ \emph{for a suitable choice of the local
bases} $\{ F_a^\sista | \ a=1,\ldots,\naq\}$, $\{ F_\alpha^\sistb |\
\alpha=1,\ldots,\nbq\}$ in $\lha$, $\lhb$. It follows that
\begin{equation}
\| \rho \ccn = \| \chrho^\rea \|_{\tr} .
\end{equation}
Therefore, the RC admits the following remarkable formulation
involving the \emph{sum of the Schmidt coefficients}:
\begin{equation}\label{RC}
\rho \ \ \mbox{separable} \ \ \Rightarrow \ \ \| \rho \ccn =
\sum_{a=1}^\dd \mu_a \leq 1.
\end{equation}
This second more conceptual formulation of the RC explains the
alternative term `computable cross norm criterion'. This formulation
also highlights the fact that the RC relies only on the Schmidt
equivalence class of a bipartite state.


\section{Entanglement and the symmetric polynomials in the Schmidt coefficients}
\label{sec:ent-bis}

The RC is based on the evaluation of a single quantity in terms of
which a necessary condition for separability (equivalently, a
sufficient condition for entanglement) can be expressed. As we have
seen, this quantity is the sum of the Schmidt coefficients of a
bipartite state. It is quite natural to wonder whether the
\emph{whole set} of the Schmidt coefficients of a bipartite state
may allow a stronger characterization of entanglement with respect
to the RC alone. Otherwise stated, it is natural to look for further
separability criteria relying only on the Schmidt equivalence class
of a state. The issue is how to extract, if possible, the relevant
information content from the Schmidt coefficients. Recently, in
ref.~{\cite{Lupo}}, it has been proposed to consider the
(elementary) \emph{symmetric polynomials} in the Schmidt
coefficients of a state $\rho\in\dens$, namely:
\begin{eqnarray}
\minor^{[1]} (|\trho|) \spa & = & \spa \sum_{a=1}^\dd \mu_a \,,
\hspace{3.2cm} \mbox{(first order symmetric polynomial)}
\nonumber\\
\minor^{[2]} (|\trho|) \spa & = & \spa \sum_{a < b} \mu_a \mu_b \,, \nonumber\\
& \dots & \\
\minor^{[l]} (|\trho|) \spa & = & \spa \sum_{ a_1<a_2< \dots < a_l
}\hspace{-0.8mm} \mu_{a_1}
\mu_{a_2} \dots \mu_{a_l} \,, \nonumber \\
& \dots & \nonumber \\
\minor^{[\dd]} (|\trho|) \spa & = & \spa \prod_{a=1}^\dd \mu_a\, .
\hspace{3.2cm} \mbox{($\dd$-th order symmetric polynomial)}
\nonumber
\end{eqnarray}
Note that we are regarding these polynomials as functions of the
absolute value $|\trho|$ of the matrix $\trho$ associated with the
state $\rho$, see~{(\ref{loo_mat})} (the specific choice of the
local bases is irrelevant). In fact, as we have seen, the Schmidt
coefficients of $\rho$ coincide with the singular values of the
matrix $\trho$; equivalently, with the eigenvalues of the positive
square matrix $|\trho|=\sqrt{\trho^\dag\hspace{0.4mm}\trho}\,$.
Therefore, the quantities
$\minor^{[1]}(|\trho|),\minor^{[2]}(|\trho|),\ldots,\minor^{[\dd]}(|\trho|)$
appear as coefficients in the characteristic polynomial of the
matrix $|\trho|$:
\begin{equation}\label{c.poly}
\chi_{|\trho|}(x) = \det (|\trho|-x\mathbb{I}) = (-x)^\dd +
\sum_{l=1}^\dd \minor^{[l]}(|\trho|)\, (- x)^{\dd-l}.
\end{equation}
In particular, the RC is based on the evaluation of an upper bound
--- on the set of separable states --- for the symmetric polynomial
of degree one $\minor^{[1]} (|\trho|)$ (i.e.\ $\minor^{[1]}
(|\trho|)\le 1$, for $\rho$ separable).

It is then natural to look for analogous upper bounds for the higher
order symmetric polynomials. By a naive reasoning, we can say that,
if the sum of the $\dd$ Schmidt coefficients equals $S$, their
product is upper bounded by $(S/\dd)^{\dd}$. Hence we have the
following necessary condition for the separability of a density
operator $\rho\in\dens$:
\begin{equation}\label{Det}
\rho \ \ \mbox{separable} \ \ \Rightarrow \ \ \minor^{[\dd]}
(|\trho|) \leq \left( \frac{1}{\dd} \right)^\dd.
\end{equation}
It is clear that this separability criterion is weaker than the RC.
In general, one can consider the symmetric polynomial of degree $l$
and obtain the following necessary conditions for the separability:
\begin{equation}\label{naive}
\rho \ \ \mbox{separable} \ \ \Rightarrow \ \ \minor^{[l]}(|\trho|)
\leq {\dd \choose l} \left( \frac{1}{\dd} \right)^{l}.
\end{equation}
A straightforward refinement of this criterion consists in taking
into account the Schmidt rank $\RR$ (recall the second assertion of
Remark~{\ref{rimarca}}) --- $1\le\RR\le\dd$ --- of the operator
$\rho$; thus, we can write the following conditions:
\begin{eqnarray}\label{Minors}
\rho \ \ \mbox{separable} \ \ \Rightarrow \ \ \minor^{[l]}(|\trho|)
\leq \left\{\hspace{-1.6mm}
\begin{array}{cc}
{\RR \choose l} \left( \frac{1}{\RR} \right)^l &
\mbox{if} \ \ l \leq \RR \\
0 & \mbox{if} \ \ l>\RR
\end{array}\right. \hspace{-1.5mm} ,
\end{eqnarray}
$l=1,\ldots ,\dd$. Of course, for $l=1$, we have the RC. For $l\neq
1$, $l\le\RR$, the necessary conditions~{(\ref{Minors})} are
consequences of the RC; hence, they are weaker separability
criteria. The existence of upper bounds, for the symmetric
polynomials of degree $l \ge 2$, stricter than the `naive' bounds
in~{(\ref{Minors})} --- in particular, stricter upper bounds that
may give rise to new separability criteria independent on the RC ---
has been investigated~{\cite{Lupo}}. Numerical tests indicate that
such stricter upper bounds indeed exist, but, till now, an
analytical estimation has not been obtained. On our opinion, this is
another interesting open problem.


\section{The symmetric polynomials in the Schmidt coefficients and the entanglement breaking quantum channels}
\label{duality}

The RC --- as well as the derived separability criteria based on the
symmetric polynomials in the Schmidt coefficients discussed in the
previous section --- can be applied to the study of quantum
channels, i.e.\ of completely positive trace-preserving (CPT) maps
by exploiting the well known correspondence between quantum channels
and quantum states (see~{\cite{MaSt,Jam,ZycBen}}).

We will consider, as above, a pair of quantum systems with carrier
Hilbert spaces $\hilba$ and $\hilbb$, and we will denote by
$\mathrm{CPT}(\hilbb,\hilba)$ the set of CPT maps from the system
$\sistb$ to system $\sista$; i.e.\ from $\lhb$ into $\lha$. Given a
CPT map $\mathcal{E} \in \mathrm{CPT}(\hilbb,\hilba)$, one can
associate with $\mathcal{E}$ a state $\rho_{\mathcal{E}} \in
\mathcal{D}(\hilba\otimes\hilbb)$ in the following {\it canonical}
way:
\begin{equation}\label{can}
\mathcal{E} \ \ \longmapsto \ \ \rho_\mathcal{E} = ( \mathcal{E}
\otimes \mathcal{I} ) (\rho_{\mathcal{I}}),
\end{equation}
with $\rho_{\mathcal{I}} = |\psi\rangle\langle\psi| \in
\mathcal{D}(\hilbb\otimes\hilbb)$ denoting a maximally entangled
(pure) state --- namely, $|\psi\rangle = \frac{1}{\sqrt{\nb}}
\sum_{\alpha=1}^\nb |\alpha\rangle\otimes|\alpha\rangle$, for some
orthonormal basis $\{ |\alpha\rangle\}_{\alpha=1}^\nb$ in $\hilbb$
--- and $\mathcal{I}$ denoting the identity in $\lhb$.

A quantum channel $\mathcal{E}\in \mathrm{CPT}(\hilbb,\hilba)$ is
said to be \emph{entanglement breaking} (EB) if the extended map
$\mathcal{E}\otimes\mathcal{I}$ transforms any state into a
separable one~{\cite{EnBr}}.
One can show --- see~{\cite{EnBr}} --- that a CPT is EB if and only
if $\mathcal{E}\otimes\mathcal{I}$ maps a maximally entangled state
into a separable one; hence, if and only if the canonically
associated state is separable.
In particular, a separability criterion also gives a necessary
conditions for a channel to be entanglement breaking.
The RC and the derived criteria based on symmetric polynomials in
the Schmidt coefficients have an interesting interpretation when
considered as criteria for EB channels.

In fact, choose local orthonormal bases $\{ F_a^\sista | \
a=1,\ldots,\naq\}$, $\{ F_\alpha^\sistb |\ \alpha=1,\ldots,\nbq\}$
in $\lha$ and $\lhb$, respectively, and consider the `matrix
elements'
\begin{equation}
\tilde{\mathcal{E}}_{a\alpha} = \tr\hspace{-0.4mm}\left(
{F_a^\sista}^\dag \hspace{0.3mm} \mathcal{E}\hspace{-0.7mm}\left(
{F_\alpha^\sistb}^\dag\right) \right).
\end{equation}
It is easy to check that the canonically associated density matrix
$\rho_\mathcal{E}$ has the following form:
\begin{equation}
\rho_\mathcal{E} = \frac{1}{\nb}\, \tilde{\mathcal{E}}_{a\alpha}\,
F_a^\sista\otimes F_\alpha^\sistb.
\end{equation}
Then, we can consider the characteristic polynomial
\begin{equation} \label{stateform}
\chi_{|\trho_{\mathcal{E}}|}(x) = \det \Big(
\frac{1}{\nb}\,|\tilde{\mathcal{E}}| - x \mathbb{I} \Big) = (-x)^\dd
+ \sum_{l=1}^\dd \minor^{[l]}(|\tilde{\mathcal{E}}|)\,
\Big(\frac{1}{\nb}\Big)^l (-x)^{\dd-l}.
\end{equation}
Relation~{(\ref{stateform})} establishes a link between the
symmetric polynomials in the singular values of the matrix
$\tilde{\mathcal{E}}$ (the eigenvalues of $|\tilde{\mathcal{E}}|$)
and the symmetric polynomials in the Schmidt coefficients of the
state $\rho_{\mathcal{E}}$. Thus, one can apply the results of the
preceding section.

Let us consider, for instance, the case where the matrix
$\tilde{\mathcal{E}}$ has full rank. Then, one obtains from
relation~{(\ref{Det})} the following necessary condition for the
channel $\mathcal{E}$ to be entanglement breaking:
\begin{equation}
\mathcal{E} \ \ \mbox{entanglement breaking} \ \ \Leftrightarrow \ \
\rho_{\mathcal{E}} \ \ \mbox{separable} \ \ \Rightarrow \ \ \det
(|\tilde{\mathcal{E}}|)\leq\left(\frac{\nb}{\dd}\right)^\dd.
\end{equation}
This result relates a geometric property of the map --- such as the
rate of contraction of volume, which equals
$\det(|\tilde{\mathcal{E}}|)$ --- to the property of being
entanglement breaking.
In general, relation~{(\ref{Minors})} implies that, if the matrix
$\tilde{\mathcal{E}}$ has rank $\RR$, we can write the following
necessary conditions for the channel $\mathcal{E}$ to be
entanglement breaking:
\begin{equation}
\mathcal{E} \ \ \mbox{entanglement breaking} \ \ \Rightarrow \ \
\minor^{[l]}(|\tilde{\mathcal{E}}|) \leq {\RR \choose l}
\left(\frac{\nb}{\RR} \right)^{l},
\end{equation}
for $l=1,\dots ,\RR$.


\section{Entanglement and the Schmidt coefficients of transformed operators}
\label{sec:ineq}

As we have seen in the preceding sections, one can fruitfully
extract some `information' from the Schmidt coefficients for
detecting entanglement. In particular, the RC relies on the sum of
the Schmidt coefficients of a bipartite state (which coincides with
the CCN of the state). A slightly more sophisticated strategy is the
following. Instead of considering the Schmidt coefficients of the
state under scrutiny itself, one may consider the Schmidt
coefficients of a `transformed operator' suitably associated with
this state. Such a transformed operator will not be, in general, a
density operator. However, one may heuristically regard this
transformation as an `entanglement highlighting procedure'
preliminary to the entanglement detection via the evaluation of the
CCN.

Let $\rho\in\dens$ be a separable state in the bipartite Hilbert
space $\mathcal{H}=\hilba\otimes\hilbb$, with a separability
decomposition of the form
\begin{equation} \label{super}
\rho = \sum_i p_i\; \rho_i^\sista \otimes \rho_i^\sistb, 
\end{equation}
where $ \sum_{i} p_i = 1$ and $p_i>0$. Let us denote by $\mara$,
$\marb$ the \emph{marginals} (reduced density operators) of $\rho$,
namely:
\begin{equation}
\mara := \trb(\rho)=\sum_i p_i\, \rho_i^\sista,\ \ \ \marb :=
\tra(\rho)=\sum_i p_i\, \rho_i^\sistb.
\end{equation}
In order to associate with $\rho$ a suitable `transformed operator',
for each of the two subsystems $\sista$ and $\sistb$ we will
consider a pair of linear or antilinear (super)operators. Precisely,
we will consider a set of operators
\begin{equation}
\Big\{\eau,\eadu\hspace{0.3mm}\colon \lha\rightarrow\lha,\
\ebu,\ebdu\hspace{0.3mm}\colon \lhb\rightarrow\lhb\Big\}
\end{equation}
that are assumed to be \emph{jointly} linear or antilinear --- i.e.\
either all linear or all antilinear --- in such a way that one can
consistently define tensor products and sums of these operators. We
will associate with $\rho$ the operator $\ope \in\lh$ defined as
\begin{equation}
\ope := \frac{1}{2} \left(\eau\otimes\ebu +
\eadu\otimes\ebdu\right)(\rho)
 + \frac{1}{2} \left(\eau\otimes\ebdu +
\eadu\otimes\ebu\right)(\mara\otimes\marb).
\end{equation}
We stress that, in general, $\ope$ is not a density operator. It is
easy to check that
\begin{eqnarray}\label{rc_dec}
\ope \spa & = & \spa \frac{1}{2}\sum_{i,j} p_i p_j\,
\Big(\eau(\rho_i^\sista)+\eadu(\rho_j^\sista)\Big)
\otimes\Big(\ebu(\rho_i^\sistb)+\ebdu(\rho_j^\sistb)\Big)
\nonumber\\
& \equiv & \spa \frac{1}{2}\sum_{i,j} p_i p_j\,
\Big(\eau(\rho_i^\sista)+\eadu(\rho_j^\sista)\Big)\otimes \abi,
\end{eqnarray}
where the symbol $\abi$ denotes repetition of the preceding term
with the substitution of the subsystem $\mathsf{A}$ with the
subsystem $\mathsf{B}$ (all the rest remaining unchanged). At this
point, one can try to obtain an estimate of the CCN of the
transformed operator $\ope$. Using relation~{(\ref{rc_dec})} and the
triangle inequality, one finds out that
\begin{eqnarray}
\left\|\ope\right\ccn \spa & = & \spa
\frac{1}{2}\left\|\mbox{$\sum_{i,j}$}\, p_i p_j\,
\Big(\eau(\rho_i^\sista)+\eadu(\rho_j^\sista)\Big)\otimes\abi\right\ccn \nonumber\\
\label{diseg1} & \le & \spa \frac{1}{2}\,\mbox{$\sum_{i,j}$}\,p_i
p_j\left\|
\Big(\eau(\rho_i^\sista)+\eadu(\rho_j^\sista)\Big)\otimes\abi\right\ccn.
\end{eqnarray}
Next, recall that the CCN $\|\cdot\ccn$ is a cross norm on $\lh$;
hence:
\begin{eqnarray}
\left\|\Big(\eau(\rho_i^\sista)+\eadu(\rho_j^\sista)\Big)\otimes
\abi\right\ccn \hspace{-3.5mm} & = & \spa \left\|\eau(\rho_i^\sista)
+ \eadu(\rho_j^\sista)\right\|_\hs\,
\hspace{-0.5mm}\left\|\ebu(\rho_i^\sistb) +
\ebdu(\rho_j^\sistb)\right\|_\hs
\nonumber\\
& = & \spa \Big(\langle
\eau(\rho_i^\sista),\eau(\rho_i^\sista)\rangle_\hs +
\langle\eadu(\rho_j^\sista),\eadu(\rho_j^\sista)\rangle_\hs
\nonumber\\
& + & \spa (\langle
\eau(\rho_i^\sista),\eadu(\rho_j^\sista)\rangle_\hs
+\coco)\Big)^{\frac{1}{2}} \abi^{\frac{1}{2}}.
\end{eqnarray}

Therefore, if we assume that --- for some $\supa,\supb \ge 0$ ---
the operators $\eau,\eadu,\ebu,\ebdu$ satisfy the conditions
\begin{equation}
\|\eau(\hsi_1^\sista)\|_\hs^2+ \|\eadu(\hsi_2^\sista)\|_\hs^2\le
2\supa,\ \ \; \|\ebu(\hsi_1^\sistb)\|_\hs^2 +
\|\ebdu(\hsi_2^\sistb)\|_\hs^2\le 2\supb,
\end{equation}
$\forall\, \hsi_1^\sista,\hsi_2^\sista\in\densa$, $\forall\,
\hsi_1^\sistb, \hsi_2^\sistb\in\densb$, we find the following
estimate:
\begin{eqnarray}
\left\|\Big(\eau(\rho_i^\sista)+\eadu(\rho_j^\sista)\Big)\otimes
\abi\right\ccn \hspace{-2.4mm} & \le &\spa 2\,
\sqrt{\left(\supa+\frac{1}{2}\Big(\langle
\eau(\rho_i^\sista),\eadu(\rho_j^\sista)\rangle_\hs
+\coco\Big)\right)} \sqrt{\abi}\,. \label{diseg2}
\end{eqnarray}
Observe that the same estimate holds true if one replaces the cross
norm $\|\cdot\ccn$ with any \emph{subcross norm} in $\lh$, i.e.\
with a norm $\|\cdot\sub\colon\lh\rightarrow\mathbb{R}^+$ such that
\begin{equation}
\|A\otimes B\sub \le \|A\|_\hs\hspace{0.3mm} \|B\|_\hs,\ \ \ \forall
A\in\lha ,\ \forall B\in\lhb.
\end{equation}

Eventually, from inequalities~{(\ref{diseg1})} and~{(\ref{diseg2})},
we obtain:
\begin{eqnarray}
\left\|\ope\right\ccn \hspace{-2.4mm} & \le & \spa
\mbox{$\sum_{i,j}$}\,\sqrt{p_i p_j
\left(\supa+\frac{1}{2}\Big(\langle
\eau(\rho_i^\sista),\eadu(\rho_j^\sista)\rangle_\hs
+\coco\Big)\right)}
\nonumber\\
& \times & \spa \sqrt{p_i p_j \left(\supb +\frac{1}{2}\Big(\langle
\ebu(\rho_i^\sistb),\ebdu(\rho_j^\sistb)\rangle_\hs
+\coco\Big)\right)}
\nonumber\\
& \le & \spa
\sqrt{\left(\supa+\frac{1}{2}\Big(\mbox{$\sum_{i,j}$}\,p_i
p_j\,\langle \eau(\rho_i^\sista),\eadu(\rho_j^\sista)\rangle_\hs
+\coco\Big)\right)} \sqrt{\abi}\, , \nonumber
\end{eqnarray}
where the second inequality above follows from the Cauchy-Schwarz
inequality. In conclusion, for every separable state $\rho\in\dens$,
the following inequality holds:
\begin{equation} \label{checkine}
\left\|\ope\right\ccn \hspace{-0.4mm} \le
\sqrt{\left(\supa+\frac{1}{2}\Big(\langle
\eau(\mara),\eadu(\mara)\rangle_\hs +\coco\Big)\right)} \
\sqrt{\abi}.
\end{equation}

The previous result can be generalized as follows. For any
$\mathsf{n}\ge 1$, consider a set of $2\mathsf{n}$ jointly linear or
antilinear operators
\begin{equation}
\Big\{\eau,\ldots,\ean\hspace{0.3mm}\colon \lha\rightarrow\lha,\
\ebu,\ldots,\ebn\hspace{0.3mm}\colon \lhb\rightarrow\lhb\Big\}.
\end{equation}
These operators allow to define a map
\begin{equation}
\map\hspace{0.3mm}\colon \hspace{0.3mm} \dens \rightarrow \lh,
\end{equation}
by setting:
\begin{equation}\label{defop}
\op := \mathsf{n}^{-1} \left(\eau\otimes\ebu +\cdots +
\ean\otimes\ebn\right)(\rho) + \mathsf{n}^{-1} \Big(\sum_{k\neq
l}\eak\otimes\ebl\Big)(\mara\otimes\marb),\ \ \ \forall \rho\in\dens
,
\end{equation}
where $\mara$ and $\marb$ are the marginals of $\rho$, namely:
$\mara:=\trb(\rho)$, $\marb:=\tra(\rho)$. We will call the operator
$\op$ the \emph{$\map$-transform} of the bipartite state $\rho$.
Then, the following result holds (see~{\cite{Aniello}}):
\begin{theorem} \label{new-new}
Let $\eau, \ldots, \ean :\lha\rightarrow\lha$, $\ebu, \ldots, \ebn
:\lhb\rightarrow\lhb$ be jointly linear or antilinear operators
that, for some $\supa,\supb \ge 0$, satisfy the following
conditions:
\begin{equation} \label{condicio1}
\|\eau(\hsi_1^\sista)\|_\hs^2+\cdots+
\|\ean(\hsi_{\ennebis}^\sista)\|_\hs^2\le \mathsf{n}\supa,\ \ \
\forall\, \hsi_1^\sista,\ldots ,\hsi_{\ennebis}^\sista\in\densa,
\end{equation}
\begin{equation} \label{condicio2}
\|\ebu(\hsi_1^\sistb)\|_\hs^2 +\cdots+
\|\ebn(\hsi_{\ennebis}^\sistb)\|_\hs^2\le \mathsf{n}\supb,\ \ \
\forall\, \hsi_1^\sistb,\ldots , \hsi_{\ennebis}^\sistb\in\densb .
\end{equation}
Then, for every separable state $\rho\in\dens$, the $\map$-transform
of $\rho$ satisfies the following inequality:
\begin{equation} \label{inefond}
\left\|\op\right\ccn \hspace{-0.4mm} \le
\sqrt{\Big(\supa+\frac{1}{\mathsf{n}}\sum_{k < l}\big(\langle
\eak(\mara),\eal(\mara)\rangle_\hs +\coco\big)\Big)\abi}.
\end{equation}
\end{theorem}

\begin{rmk} As already noticed, on the
l.h.s.\ of inequality~{(\ref{inefond})} we may replace the norm
$\|\cdot\ccn$ with any subcross norm in $\lh$. Such a replacement
may give rise to further separability criteria that would be,
however, weaker than the corresponding separability criteria
involving the CCN. Indeed, it is easy to show that
\begin{equation} \label{lastine}
\|C\sub\le\|C\ccn,\ \ \ \forall\hspace{0.3mm} C\in\lh,
\end{equation}
for every subcross norm $\|\cdot\sub$ in $\lh$.\footnote{Let
$\sum_{k=1}^n A_k\otimes B_k$ any decomposition of $C\in\lh$. Then,
by the triangle inequality and the subcross property of
$\|\cdot\sub$ we get : $\| C\sub\le \sum_{k=1}^n \|A_k\otimes
B_k\sub \le\sum_{k=1}^n \|A_k\|_\hs\, \|B_k\|_\hs$. From the
definition of the CCN inequality~{(\ref{lastine})} follows.}
\end{rmk}

Observe that Theorem~{\ref{new-new}} induces a class of separability
criteria that can be regarded as a generalization of the RC. Indeed,
for $\mathsf{n}=1$, and $\eau,\ebu$ coinciding with the identity
superoperators (so that we can set: $\supa=\supb=1$), we recover the
RC~{(\ref{RC})}. Notice also that
conditions~{(\ref{condicio1})-(\ref{condicio2})} are satisfied
--- with $\supa=\supb=1$ --- if the superoperators
$\{\eak,\ebk\}_{k=1,\ldots,\mathsf{n}}$ are such that
\begin{equation} \label{condicio3}
\|\eak(\hsi^\sista)\|_\hs\le 1,\ \ \|\ebk(\hsi^\sistb)\|_\hs\le 1,\
\ k=1,\ldots,\mathsf{n},
\end{equation}
$\forall\, \hsi^\sista\in\densa$, $\forall\, \hsi^\sistb\in\densb$.
One can assume, in particular, that they are
trace-norm-nonincreasing on positive operators (since the
Hilbert-Schmidt norm is majorized by the trace norm); for instance,
positive trace-preserving linear maps.

For $\mathsf{n}=2$ --- with the following choice of the
superoperators:
\begin{equation}
\mathfrak{E}_1^\sista = e^{\ima \theta}\hspace{0.3mm}
\mathcal{I}^\sista,\ \ \mathfrak{E}_1^\sistb =
e^{-\ima\theta}\hspace{0.3mm} \mathcal{I}^\sistb, \ \
\mathfrak{E}_2^\sista = - \mathcal{I}^\sista, \ \
\mathfrak{E}_2^\sistb = -\mathcal{I}^\sistb,\ \ \ \theta\in [0,\pi],
\end{equation}
where $\mathcal{I}^{\sista},\mathcal{I}^{\sistb}$ denote the
identity in $\lha$ and $\lhb$, respectively --- we obtain an
interesting class of separability criteria. Namely, for every
separable state $\rho\in\dens$, the following $\theta$-parametrized
family of inequalities holds:
\begin{equation}
\left\| \rho - \cos{\theta}\hspace{0.5mm} \mara\otimes\marb
\right\ccn \hspace{-0.3mm} \le
\sqrt{\left(1-\cos\theta\,\tr(\mara^2)\right)
\left(1-\cos\theta\,\tr(\marb^2)\right)}\,, \ \ \ \theta\in [0,\pi].
\end{equation}
In particular, for $\theta = \pi /2$, we re-obtain once again the
RC. For $\theta=\pi$, we recover a separability criterion recently
discovered by Zhang {\it et al.}, see ref.~{\cite{Zhang}}. This
criterion has been shown to be stronger than the RC.

Another interesting class of separability criteria contained in the
family of criteria induced by Theorem~{\ref{new-new}} are the
so-called `local filtering enhancements' of the RC,
see~{\cite{Aniello,Hyllus}}. In this case, we have $\mathsf{n}=2$
and
\begin{equation*}
\mathfrak{E}_1^\sista \colon \lha\ni A\mapsto \filta A
\hspace{0.5mm}\big(\filta\big)^\dagger \hspace{-1mm}\in\lha,\
\mathfrak{E}_1^\sistb \colon \lhb\ni B\mapsto \filtb B
\hspace{0.5mm}\big(\filtb\big)^\dagger \hspace{-1mm} \in\lhb,
\end{equation*}
\begin{equation}
\mathfrak{E}_2^\sista \colon \lha\ni A\mapsto \ima
\mathfrak{E}_1^\sista (A)\in\lha, \ \mathfrak{E}_2^\sistb \colon
\lhb\ni B\mapsto -\ima \mathfrak{E}_1^\sistb (B)\in\lhb,
\end{equation}
where $\filta,\filtb$ are suitable linear operators in $\lha$ and
$\lhb$, respectively (such that we can set: $\supa=\supb=1$); hence,
for every separable state $\rho\in\dens$,
\begin{equation}
\left\|\big(\filta\otimes\filtb\big)\, \rho\,
\big(\filta\otimes\filtb\big)^\dagger \right\ccn\hspace{-0.4mm}\le
1\, .
\end{equation}
`Optimal choices' of the the operators $\filta,\filtb$ (for a given
state $\rho$ under scrutiny) can be achieved by means of a
constructive procedure, see~{\cite{Hyllus}} and references therein.
It would be interesting to investigate the possibility of extending
such a procedure to a broader class of separability criteria induced
by inequality~{(\ref{inefond})}.


\section{Conclusions} \label{conclusions}

The detection of the entanglement of a bipartite state is a
formidable problem that is nowadays the object of intense
investigation by a broad scientific community. Among the several
separability criteria proposed so far in the literature, we think
that a prominent position --- at least from the conceptual point of
view --- is occupied by the \emph{realignment criterion} (RC),
which, as we have seen, admits two different formulations. In its
more `simple-minded' formulation~{\cite{ChenWu}} (that is, however,
of very practical use) it does not reveal immediately its profound
meaning, which is related to the \emph{computable cross norm}
(CCN)~{\cite{Rudolph,Rudolph-bis}}. The formulation of the RC in
terms of the CCN --- see Sect.~{\ref{sec:ent}} --- brings under our
attention the role played by the \emph{Schmidt coefficients} of a
bipartite state. In particular, the RC relies on the sum of the
Schmidt coefficients (which coincides with the CCN of the state).
However --- see Sect.~{\ref{sec:ent-bis}} --- there is evidence that
the whole set of the Schmidt coefficients (rather than their sum
alone) may allow a stricter characterization of entanglement with
respect to the RC. Furthermore, as shown in Sect.~{\ref{sec:ineq}},
one can adopt a more sophisticated strategy in order to extract
information about entanglement from the Schmidt coefficients. This
strategy consists in first suitably transforming a state and then
considering the Schmidt coefficients of the transformed operator.

During our brief survey on the role of the Schmidt coefficients in
the detection of entanglement, we have indicated various open
problems. First, as mentioned in Sect.~{\ref{sec:classes}}, it would
be interesting to achieve a deeper understanding of the structure of
the Schmidt equivalence classes of states; in particular, to
ascertain whether they are differentiable manifolds seems to be a
harsh mathematical problem that would deserve further attention.
Next, as observed in Sect.~{\ref{sec:ent-bis}}, it would be
desirable to extract from the Schmidt coefficients more information
about entanglement than that revealed by their sum --- i.e.\ by the
CCN of a state --- for instance, by means of suitable functions of
these coefficients (e.g., the symmetric polynomials). In other
words, it would be interesting to find further separability criteria
relying on the Schmidt equivalence classes only. Finally, another
interesting open question is the possibility of obtaining a further
refinement of the entanglement detection strategy outlined in
Sect.~{\ref{sec:ineq}}, with the aim of selecting `optimal
transformations' (with regard to the entanglement detection) for
each given state under scrutiny, as in the special case of the
`local filtering enhancements' of the RC.


\section*{Acknowledgments}
The main results of the paper were presented by the authors at the
international conference \emph{The Jubilee 40th Symposium on
Mathematical Physics -- Geometry \& Quanta} (25-28 June 2008, Torun,
Poland). They wish to thank the organizers for their very kind
hospitality.



\end{document}